\documentclass[12pt]{article}

\usepackage{amsmath}
\usepackage{amssymb}

\newcommand{\eps}{\varepsilon}
\newcommand{\ga}{\gamma}

\newcommand{\prt}{\partial}

\begin{document}

\title{Inelastic interactions between nuclei at high energies
\footnote{Zh. Eksp. Teor. Fiz. {\bf 70}, 768--784 (1976) [Sov. Phys. JETP
{\bf 43,} No.~3, 397-405 (1976)]}}

\author{V.G. Nosov$^{\dagger}$ and A.M. Kamchatnov$^{\ddagger}$\\
$^{\dagger}${\small\it Russian Research Center Kurchatov Institute, pl. Kurchatova 1,
Moscow, 123182 Russia}\\
$^{\ddagger}${\small\it Institute of Spectroscopy, Russian Academy of Sciences,
Troitsk, Moscow Region, 142190 Russia}
}

\maketitle

\begin{abstract}
A theory of nucleus-nucleus collisions has been developed for kinetic energies
substantially in excess of the binding energy. The very high pressure produced
in the compound system as a result of the fusion of the two colliding nuclei
is the reason for the subsequent hydrodynamic expansion of the nuclear medium.
The energy and angular distributions of the reaction products are investigated.
The charge distribution is also determined  in  the case  where  the  nucleon
and  ion components of the reaction  products are predominant.  A solution is
found for the expansion into vacuum of a sphere in which the initially uniformly
distributed material is initially at rest and at an ultrarelativistic temperature.
\end{abstract}

\section{Introduction}

Progress in the technology of acceleration of multiply charged ions [1] has
substantially contributed to recent developments in this important field of
research in nuclear physics.   The complexity of the colliding systems, i.e.,
the accelerated ion and the target nucleus, gives rise to a variety of possible
reaction channels specific for this category of processes.   Let $E_1$ be the
kinetic energy of the incident nucleus per nucleon.   For $E_1\sim 1\div 10$ MeV,
the nucleon binding energy in the initial systems, and the Coulomb barrier,
which impedes the approach of the two particles, may still play an appreciable
role.   This, of course, leads to an increase in the fraction of reactions
involving the transfer or capture of individual nucleons during the
interaction [2].  However, these values do not, in principle, represent the
limit of experimental possibilities, and there is a promising tendency for $E_1$
and the atomic weight of the colliding systems to increase.   The physical
picture may then be expected to undergo a substantial change, and the predominant
mechanism responsible for most of the interaction cross section turns out to
be relatively simple.

To avoid unnecessary detail with very little bearing on the essence of the
situation, we shall confine our attention to a head-on collision between two
identical nuclei and, unless stated to the contrary, we shall carry out our
analysis in the center-of-mass system (c.m.s.). Suppose that the kinetic energy
per nucleon in this system is $E_0$, the mass number of the nuclei is $A$, and
the atomic number is $Z$. For sufficiently high values of $E_0$, we can neglect
electric forces and assume that, as the nuclei approach one another,  an
overlap between the spatial distributions of the initially cold medium will
appear from a certain instant of time onward.   To obtain an approximate measure
of the strong interaction which results in this situation, let us estimate the
mean free path.   The cross section $\sigma_{nn}$ for the interaction between
the initial elementary particles, i. e., nucleons, is known from experiment
(see, for example, [3]) and is of the order of the pion Compton wavelength
$\hbar/m_{\pi}c$, whereas the role of the various constraints imposed by the
Pauli principle decreases with increasing $E_0$. Very approximately, therefore,
we have
\begin{equation}\label{1}
    \frac1{\sigma_{nn}n}\sim\frac{\hbar}{m_{\pi}c},
\end{equation}
where $n$ is the density of nucleons in nuclear matter (the possible creation
of new particles will require a more careful analysis, and in the discussion
below we shall interpret $n$ as the spatial density of baryon charge).
The mechanism and the possibility of a theoretical description of the phenomenon
are very dependent on the relative free path given by (1) and the nuclear radius
\begin{equation}\label{2}
    R=r_0A^{1/3}.
\end{equation}
It is well known that
$$
r_0\sim\hbar/m_\pi c\approx 1.4\cdot 10^{-13}\,\mathrm{cm}
$$
(this has already been used above in estimating $n$).   We thus find that,
in the case in which we are interested here\footnote{In our previous papers
[4--7] on the macroscopic treatment of apparently unrelated nuclear phenomena,
we always came across the condition $k_fR\gg1$, where $k_f$ is the
limiting momentum of the Fermi liquid quasiparticles.   It is readily
seen that the condition given by (3) reduces to a very similar criterion.}
\begin{equation}\label{3}
    1/\sigma_{nn}nR\sim A^{-1/3}\ll 1.
\end{equation}
Since the mean free path is short, the initial stage is the fusion of
the nuclei into a ``compound system". It is, however, important to emphasize
the difference between this system and the usual compound nuclei formed, say,
by nucleon capture.   In cold or relatively low-temperature nuclei there is
no appreciable pressure, and such nuclei exhibit no noticeable tendency to
expand.   On the other hand, a very high pressure is produced during the
formation of the system in which we are interested.   Thus, the most conservative
estimates, which do not take into account the compression of the medium during
fusion, show that the pressure is proportional to the total internal energy
$E = 2AE_0$.   This results in the expansion of the compound system into vacuum.
The condition given by (3) enables us to consider the second stage, i. e.,
expansion, in hydrodynamic terms.   It is not clear whether any systematic
theory would be capable of providing a detailed quantitative description of the
``fusion stage".   The essential feature is that the entropy of the system
increases from zero to some maximum value $S$.

The formulation of the problem is thus quite close to the suggestion put forward
at one time, on Fermi's initiative [8], for the description of collisions between
relativistic strongly-interacting elementary particles.
These ideas were extended further in an interesting paper by
Landau [9].  Without going into the various aspects of this complex problem,
we shall merely note some of the features of the difference between the nuclear
case and the ``elementary interaction" between two hadrons.   The initiating
interaction between the two initial particles may in itself be capable of creating
``hadron matter" in macroscopic amounts, but this entire question is, to some
extent, shrouded in doubt.   On the other hand, during the fusion of heavy
nuclei, the number of particles is known to be macroscopically large because of
the nucleons that are present right from the beginning.   This enhances the
credibility of the above thermodynamic and hydrodynamic conditions.

Strictly speaking, only the second stage of the process, i. e., the expansion stage,
will be subjected to theoretical analysis.   Let us begin with a few preliminary
remarks on the physics of the phenomena involved in this process.   The expansion
of matter into vacuum occurs with near-sonic or ultrasonic velocities, so that
viscous friction and thermal conductivity can hardly be expected to lead to an
appreciable increase in entropy. If the resulting adiabatic motion ($S =$ const)
of the continuous medium is to be treated in a simplified mechanistic way, the
internal energy of the liquid will, so to speak, play the role of potential energy.
This provides a clear physical picture of why the overall character of the motion
of the individual elements of the medium (the fluid particles) depends on the order
of magnitude of the velocities communicated to them.

In the case of inelastic collisions between nuclei, the physically interesting region
is the relatively extensive nonrelativistic region $E_0\ll m_nc^2$ ($m_n$ is the
nucleon mass), which is even more accessible to current experimental possibilities.
We shall write the nonrelativistic energy of a fluid particle in the form of the sum
$$
\frac12Mv^2+\eps,
$$
i.e., the sum of kinetic and potential energies ($M =$ const by definition and
$\eps$ is the internal energy).   In very approximate calculations, we can initially
ignore the energy of interaction with the ambient fluid (i. e., the work done by
pressure), and suppose that the velocity $v$ increases due to the reduction in
$\eps$ during the adiabatic expansion.   The latter leads to a subsequent reduction
in pressure, so that the assumption that the interaction between the fluid particles
is small will become increasingly valid.   The net result is that $Mv^2\gg\eps$, i.e.,
the fluid particles become ``freed" and execute inertial motion with $v\approx$ const.
This condition also determines the validity of the assumption that the true particles
of the medium have negligible thermal velocities (due to cooling on expansion), as
compared with the translational velocity $v$ of the fluid.   Thus, the final velocity
distribution of the particles, i.e., the reaction products, is predetermined while,
on the other hand, the hydrodynamic conditions which demand that the mean free path
is small in comparison with the linear dimensions of the entire system may still be
valid. Essentially, this stage is actually reached in a relatively short time
$t\gtrsim l/u_0$, where $l$ represents the linear dimensions of the system before
the onset of expansion, and $u_0$ is the initial velocity of sound.

The foregoing general ideas lose their validity as we enter the ultrarelativistic
region $E_0\gg m_n c^2$.   The single expression
$$
\frac\eps{(1-v^2/c^2)^{1/2}}
$$
cannot be divided into ``kinetic" and ``potential" components in an entirely natural
fashion.   We note that the idea of a ``freed" liquid particle is not altogether
consistent with the general character of relativistic relationships.   It is clear,
for example, that the reduction in $\eps$ should be compensated by a reduction in
the denominator.   The fluid continues to accelerate and, in reality, the pressure
[for which at ultrarelativistic temperatures one usually employs the equation of
state given by (30)] remains effective.   The only process capable of terminating
the reduction in $\eps$ during expansion, and of stabilizing the velocity, is the
formation of individual particles in the hadron matter, the rest masses of which
begin to dominate all the contributions to the internal energy.   Here again we
return to the situation where the energy and angular distribution of the reaction
products are predetermined and, correspondingly, the equation of state for the
medium changes and departs from (30).   In the opinion of Pomeranchuk [10] and
Landau [9,11], this occurs at temperatures $T\sim m_\pi c^2$.

\section{Collisions of nonrelativistic nuclei}

We shall suppose below that the change in the internal state of the medium during
expansion is described by the Poisson adiabatic curve [12]:
\begin{equation}\label{4}
    pV^\gamma=\mathrm{const}.
\end{equation}
If we recall that $dE = -pdV$ and integrate, we can write the basic relationships
in the following form, which is particularly convenient for subsequent calculations:
\begin{equation}\label{5}
    \ga=\frac{2\nu+3}{2\nu+1},\quad w=\frac{2\nu+1}2u^2,\quad dw=(2\nu+1)udu.
\end{equation}
In simple cases, the parameter $\ga$ is the ratio of specific heats, but this is
not essential; $w$ is the enthalpy per unit mass.   Moreover, for adiabatic
(isentropic) flow
\begin{equation}\label{6}
    s\propto n\propto\rho\propto u^{2\nu+1},
\end{equation}
where $\rho$ is the density of the spatial mass distribution and $s$ is the entropy
per unit volume.   For the so-called simple (self-similar) rarefaction wave, we have
\begin{equation}\label{7}
    u+\frac{v}{2\nu+1}=u_0,
\end{equation}
where $u$ is the local velocity of sound (see, for example, [13]) and
\begin{equation}\label{8}
    v_{max}=(2\nu+1)u_0
\end{equation}
is the limiting value of the velocity of free expansion of the medium which
is initially at rest in vacuum.

To obtain an estimate for the preliminary compression of nuclear matter, we shall
suppose that the fusion of nuclei occurs gradually.   Initially, in the region
of space where the two media have come into contact, the liquid undergoes intensive
``boiling" but, outside this region, it remains cold.   Since, prior to collision,
the product of the nucleon momentum by its velocity is $2E_0$, we find that the
momentum transported out of the ambient space through unit area on the separation
boundary per unit time is
\begin{equation}\label{9}
    p=2n_0E_0,
\end{equation}
where $n_0$ is the usual equilibrium density of the baryon charge at zero temperature.
The momentum transfer specified by (9) is obviously equivalent to a pressure $p$.
After the medium has been brought to the boil, the pressure in the medium is
approximately given by
\begin{equation}\label{10}
    p=\frac23nE_0,
\end{equation}
which is the equation of state for an ideal gas\footnote{This means that we are
neglecting the potential energy of the interaction between the nucleons.
The assumption that an ideal gas is produced seems, at first sight, to be somewhat
drastic.   Nevertheless, there are reasons to suppose that it does,  in fact,
lead to a reasonable description of the main features of the phenomenon.
It is clear from the foregoing that the resulting particle-energy distribution
essentially reflects the hydrodynamic character of the process,  but is not too
sensitive to the particular choice of the adiabatic curve. It is also important
to remember that the contribution of the interaction energy rapidly decreases
during the expansion process.}.
Assuming that mechanical equilibrium is established more rapidly than thermal
equilibrium in the neighborhood of the separation boundary, we can equate the
expressions given by (9) and (10).   This yields
\begin{equation}\label{11}
    n/n_0\approx 3
\end{equation}
prior to the onset of free expansion.   We emphasize that the result given by
(11) is insufficient to determine both the longitudinal and transverse size of
the figure at the very beginning of the hydrodynamic stage.   The medium may
undergo some flow in the transverse plane which is perpendicular to the $x$ axis
during the fusion of the nuclei, and this is not impeded by external pressure.
More accurate estimates of the radial size $L >R$ reached in this direction are
difficult because of the highly non-equilibrium character of the fusion
stage (see also the Introduction).

Let us now consider the adiabatic stage of the expansion process.   The initial
configuration can be schematically represented by a disk of thickness $2l$.
It is natural to assume that
\begin{equation}\label{12}
    l\ll L.
\end{equation}
In the first approximation, therefore, the hydrodynamic flow can be looked upon
as one-dimensional.   The symmetry of the problem enables us to confine our
attention to the region $x>0$.   In addition to the coordinate measured from
the center of symmetry, it will occasionally be useful to use the variable
$x' =x - l$.   The edge of the distribution of matter moves forward with the
velocity given by (8).   As long as $t<l/u_0$, the situation is no different
from the solution of the well-known problem on the expansion of a half-space
into vacuum. Against the flow, we have the propagation of a simple wave up to
the ``weak discontinuity" $x' = -u_0t$ (i. e., the point at which the sonic
signal reaches at this time; see, for example, [13]).   When $t>l/u_0$, the weak
discontinuity moves in the positive direction of the $x$ axis, and the relative
size of the region occupied by the simple wave decreases rapidly.   The space
on the other side of the weak discontinuity, where the appropriate value of the
so-called general integral of hydrodynamic equations\footnote{In the model example
corresponding to $\nu =0$, the ``joining" of the general integral to the
simple wave is readily achieved exactly and in an explicit form for any
time $t>l/u_0$.   The fraction of energy and entropy which is asymptotically
taken up by the simple wave turns out to be of the order of $l/u_0t \ll1$.
Similar estimates are characteristic for other values of $\nu$.   It must not,
however, be supposed that the fact that the simple wave is negligible for
large times $t$ is a universal feature of all hydrodynamic problems involving
the free expansion of material into vacuum.   In the ultrarelativistic case,
the fluid is rapidly accelerated and tends to the limiting (light) velocity
so that, in general, a considerable fraction of the total energy and total
entropy is concentrated in the simple wave.   A specific example of this is
the problem solved in the Appendix.} is reached, begins to
play the dominant role.   In principle, a general analytic expression can be
obtained for it for integral values of $\nu$ [13,14].

In the case of an ideal gas of elementary particles, we have $\ga = 5/3$ and
$\nu = 1$. The corresponding general solution can be written in the form
\begin{equation}\label{13}
    \chi(w,v)=\frac1u\left\{F_1\left(u+\frac{v}3\right)+
    F_2\left(u-\frac{v}3\right)\right\},
\end{equation}
where $F_1$ and $F_2$ are some arbitrary functions.   The ``velocity
potential" can be used for the implicit determination of the required functions
$w(x', t)$ and $v(x', t)$ from the formulas
\begin{equation}\label{14}
    t=\frac{\prt\chi}{\prt w},\quad x'=v\frac{\prt\chi}{\prt w}
    -\frac{\prt\chi}{\prt v}
\end{equation}
(see, for example, [13]).   By satisfying the boundary conditions
both at $x' = - l$ $(x = 0)$, at which the fluid is at rest, and at the point
of contact with the simple wave (7), we finally obtain
\begin{equation}\label{15}
    \chi=\frac32\frac{l}u\left\{\left(u+\frac{v}3\right)^2-u_0^2\right\}.
\end{equation}
By substituting in (14) [see also (5)], we immediately return to the physically
most interesting time $t\gg l/u_0$ (in which case, $u\ll u_0$, where $u_0$ is
the initial velocity of sound in the originally resting medium):
\begin{equation}\label{16}
    v=\frac{x}t,\quad \rho\propto u^3=\frac{l}{2t}\left(u_0^2-\frac{v^2}9\right).
\end{equation}
The velocity field given by (16) corresponds to the inertial motion of the fluid
particles, and the velocity distribution of the masses remains unaltered (see
also the preliminary remarks in the Introduction).   In fact, at any time
\begin{equation}\label{17}
    \rho dx\propto\rho dv\propto(u_0^2-v^2/9)dv.
\end{equation}
If we now transform to the new variable defined by $v\propto\sqrt{\eps},$
$dv\propto d\eps/\sqrt{\eps}$ and normalize the expression $W(\eps)d\eps
\propto\rho dx$ to the unit integral between $0$ and $\eps_{max}$, we obtain the
following expression for the energy distribution of the reaction products, i.e.,
nucleons, in the center-of-mass system (see Fig.~1):
\begin{equation}\label{18}
    W(\eps)d\eps=\frac{3/4}{(\eps_{max})^{3/2}}(\eps_{max}-\eps)
    \frac{d\eps}{\sqrt{\eps}}.
\end{equation}
The presence of the cutoff point $\eps=\eps_{max}$ in (18) is a consequence of
the hydrodynamic character of the expansion stage.   The energy $\bar{\eps}$
averaged over the entire spectrum is given by
\begin{equation}\label{19}
    \eps_{max}=5\bar{\eps}=5E_0
\end{equation}
($\bar{\eps}=E_0$ follows from energy conservation).

When (12) is satisfied, the angular distribution of the nucleons is confined to
the forward and backward directions.   It cannot be calculated in a closed form,
and we shall therefore confine our attention to an estimate of the characteristic
angle $\theta\approx v_y/v$, where $v_y$ is the transverse component of the
fluid-particle velocity.   Its total acceleration $d\mathbf{v}/dt$ is given by the
Euler equation, the transverse component of which is
\begin{equation}\label{20}
    \frac{dv_y}{dt}=-\frac{\prt p/\prt y}\rho\sim\frac{u^2}L\sim
    \frac{l^{2/3}u_0^{4/3}}{Lt^{3/2}}.
\end{equation}
The solution given by (16) is used here for approximate purposes and is valid for
$u_0t \lesssim L$, after which expansion enters the three-dimensional phase.
Integrating up to the above limit,  and recalling that $v\sim u_0$, we find that
\begin{equation}\label{21}
    \theta\sim (l/L)^{2/3}.
\end{equation}

Transforming to the laboratory system, in which one of the nuclei was at rest
prior to collision, we obviously obtain $E_1 =4E_0 = 4\bar{\eps}$ for the primary
energy per nucleon.   For most particles, the observed angle $\vartheta$ between their
momenta and the collision axis has the same order of magnitude, i.e.,
$\vartheta\sim\theta\ll 1$.   If we apply the Galilean transformation to (18),
we can readily show that
\begin{equation}\label{22}
    W(\eps_1)d\eps_1=\frac{3/8}{(5\bar{\eps})^{3/2}}\left[5\bar{\eps}-
    \left(\sqrt{\eps_1}\pm\sqrt{\bar{\eps}}\right)^2\right]
    \frac{d\eps_1}{\sqrt{\eps_1}},
\end{equation}
where $\eps_1$ is the laboratory nucleon energy, and the upper and lower signs
refer to particles travelling in the forward and backward directions in this
frame, respectively.   The distribution given by (22) is normalized to a unit
total integral evaluated over both regions, and the corresponding branches of it
are shown in Fig.~1. We note that, when $\eps_1\lesssim E_1\theta^2$, the
angular distribution of the nucleons becomes broad, filling the entire solid
angle (in the laboratory system).

The foregoing discussion was, in fact, confined to the case $E_0\lesssim m_\pi c^2$.
When the inequality
\begin{equation}\label{23}
    m_\pi c^2\ll E_0\ll m_nc^2
\end{equation}
is satisfied, the situation is modified somewhat because
of the creation of a large number of relativistic pions during the fusion of
the nuclei.   In the region defined by (23), most of the mass is carried by the
nucleons.   On the other hand, the internal energy (less the nucleon rest mass,
as is usually assumed in non-relativistic theory) resides mainly in the meson degrees
of freedom, and these particles are also largely responsible for the pressure in
the medium.   By analogy with black-body radiation [12] [see also the next section
and, in particular, the equation of state given by (30)], the pressure will be
approximately specified by the equation
\begin{equation}\label{24}
    p=\frac13nE_0.
\end{equation}
Equating the pressure given by (24) to the external pressure given by (9),
which, during the fusion stage, describes the cold part of the medium for a
certain interval of time, we find that
\begin{equation}\label{25}
    n/n_0\approx 6.
\end{equation}
The increase in the preliminary compression as compared with (11) suggests that
the validity of (12) may improve \footnote{Nevertheless, the narrowness
of the region in which (23) is valid is a serious defect of the theory
applicable to it.   The condition given by (23) may not be sufficient
because the nucleon and pion rest masses are not,  in reality, very
different from one another.}.

Black-body radiation and other similar ultrarelativistic modifications of matter
correspond to $\ga = 4/3$ and $\nu = 5/2$.   In principle, for fractional values
of $\nu$, there is no closed general analytic solution of the equations of
one-dimensional hydrodynamics that are analogous to (13) and (14).   However,
for large times $t$, the asymptotic behavior of the form given by (16) can readily
be generalized to fractional values of $\nu$:
\begin{equation}\label{26}
    \begin{split}
    v&=x/t,\qquad t\gg l/u_0,\\
    \rho\propto u^{2\nu+1}&=\frac{\Gamma(2\nu+1)}{2^{2\nu}[\Gamma(\nu+1)]^2}
    \frac{l}t\left[u_0^2-\frac{v^2}{(2\nu+1)^2}\right]^\nu.
    \end{split}
\end{equation}
If we use this expression to determine the energy distribution of the particles
in the center-of-mass system, we have for $\nu = 5/2$,
\begin{equation}\label{27}
    W(\eps)d\eps=\frac{16/5\pi}{(\eps_{max})^3}(\eps_{max}-\eps)^{5/2}
    \frac{d\eps}{\sqrt{\eps}},\quad \eps_{max}=8\bar{\eps},
\end{equation}
which is valid for any reaction products with nucleons and pions predominating.
The quantity $\eps_{max}$ is proportional to the mass of the particles with which
we are concerned.   For example, $\eps_{n,max}/\eps_{\pi,max}=m_n/m_\pi$.
Because the conditions for the validity of the theory are unfavorable (see the
last footnote), the equation $\bar{\eps}_n\cong E_0$ is, in fact, satisfied only
approximately.   Finally,  if we estimate the transverse forces in the Euler
equation by analogy with the derivation of (21) from (20) and (16), we get the
expression
\begin{equation}\label{28}
    \theta\sim(l/L)^{1/3}
\end{equation}
for the effective angle at which the particles are emitted in the center-of-mass
system.   We shall not consider here the kinematics of the transformation to the
laboratory system, since it is analogous to that discussed above for the case
$E_0\lesssim m_\pi c^2$.

\section{Collisions of ultrarelativistic nuclei}

When
\begin{equation}\label{29}
    E_0\gg m_n c^2
\end{equation}
we must use the method of relativistic hydrodynamics [13,\,9,\,11].
We begin by writing down the basic thermodynamic relationships.
In the spirit of the Landau idea [9,11] on the nature and the probable form of
the equation of state for hadron matter at ultrarelativistic temperatures
$T\gg m_\pi c^2$, we assume that
\begin{equation}\label{30}
    p=\frac{e}3,\quad e=ks^{4/3},\quad T=\frac{de}{ds}=\frac43ks^{1/3}.
\end{equation}
In these expressions, $e$ is the energy per unit proper volume of the liquid
particle in its rest system, $s$ is the entropy per unit proper volume, $p$ is the
pressure, and $T$ the temperature.   This yields the following constant value
for the velocity of sound:
\begin{equation}\label{31}
    u=c/\sqrt{3}.
\end{equation}
The numerical value of $k$ cannot be established by purely deductive means.
Dimensional considerations suggest that
\begin{equation}\label{32}
    k\sim\hbar c.
\end{equation}
The volume of the compound system produced as a result of the fusion of
the original nucleus is given by
\begin{equation}\label{33}
    V\sim R^3\frac{m_nc^2}{E_0},
\end{equation}
where $R$ is the nuclear radius and the factor $m_nc^2/E_0$ appears as a result
of the Lorentz compression:
\begin{equation}\label{34}
    \frac{n}{n_0}\sim\frac{E_0}{m_nc^2}.
\end{equation}
Recalling also the expression given by (2), we can readily show that the
temperature and entropy of the system at the time preceding the onset of
adiabatic expansion are given by
\begin{equation}\label{35}
    T_0\sim m_\pi c^2\left(\frac{m_n}{m_\pi}\right)^{1/4}
    \sqrt{\frac{E_0}{m_nc^2}},\quad
    S\sim A\left(\frac{m_n}{m_\pi}\right)^{3/4}
    \sqrt{\frac{E_0}{m_nc^2}}.
\end{equation}

Let us now consider the hydrodynamic stage.   Because of the geometry of the
initial configuration, this stage has the character of one-dimensional flow
over a certain interval of time.   However, analysis shows that the increase
in the influence of transverse forces gradually leads to the isotropization
of the flow and its rapid transformation into the three-dimensional phase
\footnote{This phenomenon was considered qualitatively in the papers [9,11].
Landau called it ``lateral" or ``conical" expansion.   He used conservation laws
to predict a time dependence of the main quantities, which is confirmed by the
rigorous formula (40); see below for further details.}.
As a result, the liquid is so rapidly accelerated that it becomes concentrated
largely at finite distances from the surface which is expanding with the
velocity of light. One way of describing this is to say that a ``cavity," i.e.,
a region of sharply reduced density, is produced inside the spatial distribution
of matter with this peculiar geometry.   We shall first describe this isotropic
part of the process and will return to the influence of the initial conditions later.

It is well known that the equations of relativistic hydrodynamics are contained
in the differential conservation laws
\begin{equation}\label{36}
    \frac{\prt T^{ik}}{\prt x^k}=0,
\end{equation}
where $T^{ik}$ is the energy-momentum tensor of the medium [9,\,11,\,13,\,15].
It will be convenient to use the system of units in which $c = 1$ and adopt (30)
to simplify all the expressions to the case of spherical symmetry.
Instead of the radial distance $r$, we shall use the independent variable
\begin{equation}\label{37}
    \xi=t-r
\end{equation}
and expand into a series in powers of the reciprocal of the relativistic 4-velocity
\begin{equation}\label{38}
    \ga=\frac1{\sqrt{1-v^2}}\gg 1,
\end{equation}
retaining only the first two terms ($t/\xi\sim\ga$, as we shall soon show).
In terms of the new variables, we then obtain the following set of equations for
the ultrarelativistic flow:
\begin{equation}\label{39}
    \begin{split}
    &\ga^2\frac{\prt s}{\prt t}+2\frac1t\ga^2s+2\frac{\xi}{t^2}\ga^2s+
    \frac12\frac{\prt s}{\prt\xi}+\frac12\left[\frac{\prt\ga^2}{\prt t}-
    \frac1\ga \frac{\prt\ga}{\prt\xi}\right]s=0,\\
    &\ga^2\frac{\prt s}{\prt t}-\frac12\frac{\prt s}{\prt\xi}+
    \frac32\left[\frac{\prt\ga^2}{\prt t}+
    \frac1\ga \frac{\prt\ga}{\prt\xi}\right]s=0.
    \end{split}
\end{equation}
It is readily verified that all the requirements are satisfied by the following
very simple solution\footnote{From the more formal point of view, the simplicity
of this solution and the complexity of the one-dimensional Landau-Khalatnikov
solution [9,11,16] are probably connected with the three-dimensional
character of real physical space.   Both the hydrodynamic equations (39)
and the equation of state (30), which is taken into account in their
derivation (this also implicitly assumes the three-dimensional character
of space), correspond to this nature of physical space.   In this sense,
complete concordance of the equations of hydrodynamics and thermodynamics
in the ``one-dimensional world" also results in an exceedingly simple solution.
We shall not reproduce this solution here and merely note the following:
when the one-dimensional analog of the thermodynamic relationships given by
(30) is used, the equations of hydrodynamics turn out to be strictly linear
and can be readily solved in general form.   The solution is some explicit
function of the initial conditions.}
\begin{equation}\label{40}
    s=\frac{C}{t^3\xi^3},\quad \ga^2=\frac12\frac{t^2}{\xi^2}.
\end{equation}
The singularity at £$\xi=0$ reflects, formally, the inability of matter to
propagate with velocities in excess of the velocity of light and this is,
of course, the basic feature of the equations of relativistic hydrodynamics.
In point of fact, the value of the general integral of these equations given
by (40) is valid only up to a certain $\xi_0>0$.   The sphere $\xi=\xi_0$ is
a surface of weak discontinuity.   Integrating the differential equation of
its motion [it can be set up with the aid of the relativistic law of addition
of velocities; relative to the fluid, the weak discontinuity always propagates
with the velocity of light which, in this case, is $1/\sqrt{3}$, see (31)],
we can verify that $\xi\to\mathrm{const}$ when
\begin{equation}\label{41}
    t\gg \xi_0.
\end{equation}

To achieve a more specific physical interpretation of $\xi_0$, let us consider
the conservation laws. The spatial energy density is the time component
\begin{equation}\label{42}
    T_{00}=(e+p)\ga^2-p\simeq \tfrac43 e\ga^2
\end{equation}
of the energy-momentum tensor [9,13,15].   Moreover, the total entropy $S$
is a constant in the case of adiabatic flow.   Using (30) and (40), we obtain
\begin{equation}\label{43}
    \begin{split}
    E&=\int\frac43 ks^{4/3}\ga^2d\mathbf{r}=\frac43\cdot 4\pi t^2k\frac{C^{4/3}}
    {t^4}\frac{t^2}2\int_{\xi_0}^\infty\frac{d\xi}{\xi^6}=
    \frac8{15}\pi k\frac{C^{4/3}}{\xi_0^5},\\
    S&=\int s\ga\mathbf{r}=4\pi t^2\frac{C}{t^3}\frac{t}{\sqrt{2}}
    \int_{\xi_0}^\infty\frac{d\xi}{\xi^4}=\frac{2^{3/2}}3\pi\frac{C}{\xi_0^3},
    \end{split}
\end{equation}
which enable us to calculate $\xi_0$ and the arbitrary constant $C$:
\begin{equation}\label{44}
    \begin{split}
    \xi_0&=\frac25\left(\frac3\pi\right)^{1/3}k\frac{S^{4/3}}{E}=
    \frac25\left(\frac3\pi\right)^{1/3} V^{1/3},\\
    C&=\frac{2^{3/2}}{125}\left(\frac3\pi\right)^{2}k^3\frac{S^5}{E^3}=
    \frac{2^{3/2}}{125}\left(\frac3\pi\right)^{2} SV.
    \end{split}
\end{equation}
Consequently [see (33), and we return to ordinary units]
\begin{equation}\label{45}
    \xi_0\sim V^{1/3}\sim R(m_nc^2/E_0)^{1/3}.
\end{equation}
When the temperature is reduced to $T\sim m_\pi c^2$, the individual particles
are finally formed, and the relativistic acceleration mechanism ceases to operate
(see also the preliminary remarks at the end of the Introduction).
To estimate the corresponding time $t$, let us return to (40).   It is clear
that the volume in which the medium is concentrated is $\sim(ct)^2\xi_0$.
Moreover, $\ga\sim ct/\xi_0$. We may therefore conclude that the order of
magnitude of the ``proper volume" is $(ct)^2\xi_0\ga\sim(ct)^3$.
In this volume, the above temperature corresponds to pion separations of the
order of their Compton wavelength. Thus,
\begin{equation}\label{46}
    ct\sim \frac{\hbar}{m_\pi c}N_\pi^{1/3}.
\end{equation}
We note that this result is similar to the well-known formula given by (2).
Since $N_\pi\gg A$, comparison with (45) shows that the inequality given by (41)
is clearly satisfied.   The energy spectrum of the particles must be judged from
the entropy distribution (see [9,11]).   Its observed
density is $s\ga$ and is determined by (40).   Therefore,
\begin{equation}\label{47}
    W(\eps)d\eps\propto s\ga d\xi=s\ga\frac{d\xi}{d\ga}d\ga\propto\ga^2d\ga.
\end{equation}
This distribution cuts off sharply at $\ga=\ga_{max}$, which corresponds to
$\xi=\xi_0$ at time given by (46).   It is readily
seen that
\begin{equation}\label{48}
    \ga_{max}\sim\left(\frac{m_n}{m_\pi}\right)^{1/4}\left(\frac{E_0}{m_nc^2}
    \right)^{1/2}.
\end{equation}

The particle-energy distribution in the center-of-mass system, normalized to unity,
assumes the form
\begin{equation}\label{49}
    W(\eps)d\eps=\frac3{(\eps_{max})^3}\,\eps^2d\eps,\quad \bar{\eps}=\frac34\eps_{max},
    \quad \eps<\eps_{max}=mc^2\ga_{max}.
\end{equation}
In this expression, $m$ is the rest mass of the particular type of particles with
which we are concerned.

Let us now briefly consider the kinematics of the transformation to the laboratory system.
Elementary relativistic transformation yields:
$$
E_1\cong2\frac{E_0^2}{m_nc^2}\gg E_0,
$$
where $E_1$ is the primary laboratory energy per nucleon in the bombarding nucleus.
The particle energy and angular distribution can be found without great difficulty,
but the process is laborious.   Integrating it with respect to one of the variables,
we find that
\begin{equation}\label{50}
    \begin{split}
    W(\vartheta)do&=\frac{(m_nc^2/E_0)^2}{[\vartheta^2+(m_nc^2/E_0)^2]^2}
    \frac{do}{\pi},\\
    W(\eps_1)d\eps_1&=\frac{3/2}{\eps_{1,max}}\left[1-\left(\frac{\eps_1}
    {\eps_{1,max}}\right)^2\right]d\eps_1,\quad \eps_{1,max}=\frac{E_1}{E_0}
    \eps_{max},
    \end{split}
\end{equation}
for the angular and energy distributions, respectively. The quantity $do=2\pi
\vartheta d\vartheta$ is the solid-angle element and, as can be seen,
$\overline{\theta}\sim m_nc^2/E_0\ll 1$, i.e., the angular distribution is confined
to forward directions in the laboratory system.   We emphasize these simple
distributions are valid for the great majority of particles but, strictly speaking,
not for all of them.   As a matter of fact, the laboratory energy has the lower bound
$$
\eps_{1,min}=\frac12\frac{m}{m_n}\frac{E_0}{\ga_{max}}\ll \eps_{1,max},
$$
which is relatively low but still ultrarelativistic.   In the ``soft" part of the
spectrum adjacent to $\eps_{1,min}$, the particles are emitted at relatively large
angles right up to the maximum possible
$$
\vartheta_{max}=\frac{m_nc^2}{E_0}\ga_{max},\quad 1\gg\vartheta_{max}\gg\vartheta.
$$
These details of the ultrarelativistic distributions are illustrated in Fig.~2.

One further remark must be introduced in connection with the foregoing.   The general
principles of solution of this kind of hydrodynamic problem would appear to enable us
to say that the region of space $\xi<\xi_0$ cannot be absolutely ``empty."   It should
contain the simple wave which is in direct contact with vacuum.   Since in the equation
of state given by (30) the edge of the distribution of matter (strictly speaking, it,
too, is a weak discontinuity) always moves with the velocity of light, the radial size
$\zeta_0$ of the simple wave will also remain constant in the ultrarelativistic limit
which we have considered.   To estimate it, therefore, we must return to
an earlier stage in the expansion process.

The one-dimensional Landau-Khalatnikov theory [9,11,16] is valid for $ct\ll\xi_0$.
The initial configuration was characterized by the longitudinal size $l\sim Rm_nc^2/E_0$
[see also (33)].   When $ct>l\sqrt{3}$, both weak discontinuities move in the same,
positive, direction.   It is readily shown that, in the one-dimensional relativistic
simple wave (self-similar, see, for example, [13]), we have
\begin{equation}\label{51}
    \zeta_0\propto t^\lambda,\quad \lambda=\left(\frac{\sqrt{3}-1}{\sqrt{3}+1}
    \right)^2=7-4\sqrt{3}\approx 0.07,
\end{equation}
which describes the distance $\zeta_0$ between the weak discontinuities as a function
of time.   Thus,
\begin{equation}\label{52}
    \zeta_0\sim l\left(\frac{ct}l\right)^\lambda\sim l\sim R\frac{m_nc^2}{E_0},
\end{equation}
if, for the purpose of very approximate calculations, we neglect the effect of the
small exponent, and take into account the short duration of the entire one-dimensional
phase of the expansion process.   Comparison with (45) then yields
\begin{equation}\label{53}
    \zeta_0\ll\xi_0.
\end{equation}
We may, therefore, neglect the contributions of the energy and entropy of the simple
wave, and this was taken into account in the derivation of the formulas considered below.

\section{Charge distribution of reaction products}

The fact that the individual particles (hadrons) have certain discrete quantum numbers,
i.e., different ``charges," enables us to derive a number of interesting relationships.

The equilibrium character of the resulting electric charge distribution is clear even
from (46).   Immediately after the formation of the individual hadrons, the free path
$\sim\hbar/m_{\pi}c$ is still small in comparison with the linear dimensions of the
entire system, so that hydrodynamics and thermodynamics remain valid, as before, for
an appreciable length of time even after transition to the region $T \ll m_{\pi}c^2$
in which we have a Boltzmann gas with a practically constant number of particles
\footnote{Transition to the Boltzmann region $T\ll m_{\pi}c^2$ is accompanied by the
strong suppression of pion annihilation processes because, as the density falls,
the role of triple (and higher order) collisions falls rapidly to zero.}.
Under these conditions, elastic interactions between the particles, including
charge-transfer processes, are sufficiently effective.

We shall base our analysis on the principle of isotopic invariance (see, for example,
[17-19]).   When nuclei with the same number of protons and neutrons coalesce, the
initial state is completely isotropic in isotopic space, with all the ensuing
consequences.   In particular, all pions ($\pi^+,\,\pi^0,\,\pi^-$) are then created in
equal numbers.   However, in practice, sufficiently heavy nuclei have a neutron excess
$A-2Z$.   Using the analogy with thermodynamics, and the statistics of rotating bodies
[12], we can adhere to the point of view that, in equilibrium, a fluid particle rotates
as a whole in isotopic space with angular velocity $\Omega$.   The Boltzmann
distribution then contains the factor $\exp\{\hbar\Omega\tau/T\}$, which includes the
component $\tau$ of the particle isospin along the rotation axis.   Consequently,
\begin{equation}\label{54}
    \begin{split}
    &\frac{N_p}{N_n}=\frac{N_{\pi^+}}{N_{\pi^0}}=\frac{N_{\pi^0}}{N_{\pi^-}}=
    \exp\left(\frac{\hbar\Omega}T\right)\simeq 1+\frac{\hbar\Omega}T,\\
    &N_{\pi^+}=\frac{M_\pi}3\left(1+\frac{\hbar\Omega}T\right),\quad
    N_{\pi^0}=\frac{N_\pi}3,\quad
    N_{\pi^-}=\frac{M_\pi}3\left(1-\frac{\hbar\Omega}T\right),
    \end{split}
\end{equation}
where $N_p$ is the number of protons and $N_n$ is the number of neutrons among the
reaction products, and the other subscripts refer to pions of the appropriate type.

The validity of relationships such as those given by (54) does not depend on the
presence of other particles. Let us suppose now that antibaryons can be practically
neglected, and baryons are represented only by protons and neutrons.   Conservation
of the baryon charge $2A$ of the entire system then yields
\begin{equation}\label{55}
    N_p=A\left(1+\frac{\hbar\Omega}{2T}\right),\quad
    N_n=A\left(1-\frac{\hbar\Omega}{2T}\right).
\end{equation}
Let us now apply the conservation of electric charge $N_p+N_{\pi^+}-N_{\pi^-}=2Z$.
We have
\begin{equation}\label{56}
    \frac{\hbar\Omega}T=-2\frac{A-2Z}{A+(4/3)N_\pi},\quad
    \frac{A-N_p}{A-2Z}=\frac1{1+(4/3)(M_\pi/A)}.
\end{equation}
Thus, after the reaction, the neutron excess $A-N_p$ decreases in comparison with
its original value $A-2Z$, and hence the pion fraction contains more negative pions
than positive pions.

We note that the above formulas are even more valid for $E_0\lesssim m_nc^2$ when,
roughly speaking, there is not enough energy for antinucleon creation.   Even in the
absence of pions, the relative neutron excess is not large enough to enable us to
assume that $\hbar\Omega/T\ll1$, as above.   For the region defined by (23), we can
readily show that, very approximately,
\begin{equation}\label{57}
    N_\pi\sim S\sim\left(\frac{E_0}{m_\pi c^2}\right)^{3/4}A.
\end{equation}
In the ultrarelativistic limit $E_0\gg m_nc^2$, the situation can, at least in
principle, become modified by the creation of baryon pairs,   However, under these
conditions, since $N_\pi\sim S$ (see (35) and [9,11]), the neutron excess in the
nucleon fraction is negligible compared with the initial excess.

\section{Discussion}

Let us now briefly review the conclusions of the theory of collisions between energetic
nuclei, which refer to the energies of the individual particles after interaction.
Their mean value is, as a rule, of the order of the temperature $T_0$ of the resulting
compound system. However, the shape of the energy spectra of the reaction products
does not in itself exclude the possibility
that the mechanism may be interpretable as purely thermal and ``evaporative."
It is difficult to imagine, for example, that the restriction on the maximum energy of
the emitted particle is due to anything other than the hydrodynamic character of the
expansion of the compound system.   There is a particularly sharp jump in the
distribution function at $\eps=\eps_{max}$ in the ultrarelativistic limit (see (49)
and the explanation in text). When we refer to the nonrelativistic case
$E_0\ll m_nc^2$, we must also emphasize the shape $d\eps/\sqrt{\eps}$ of the soft part
of the spectrum, which is totally uncharacteristic for particle-evaporation processes
in the case of the usual compound nucleus.   When the necessary experimental data become
available, therefore, one would hope to be able to achieve a sufficiently reliable
identification of the hydrodynamic mechanism discussed in the present paper.

We must now briefly consider the specific features of collisions that are not of the
head-on type.   The compound system whose evolution is described by the above theory
arises in the region of space where the colliding nuclei overlap.   Those parts of
the nuclei which do not overlap remain as relatively cold fragments, in effect,
truncated on collision.   They largely continue to execute inertial motion with energy
$E_0$ per nucleon.   Subsequently, the shape of a fragment in its rest system tends
to an equilibrium, and the oscillations of the surface become transformed into heat.
The final temperature reached in the course of this process is probably a slowly-varying
function of the primary energy $E_0$ and is low.   Consequently, the velocity of the
nucleons evaporated from the fragment is also small in comparison with its translational
velocity as a whole.   The nucleons evaporated by this mechanism should therefore
produce an additional monochromatic peak at $\eps\simeq E_0$ in the energy spectrum
(in the center-of-mass system).   This interesting feature of the phenomenon suggests
that the experimental energy distributions should be even more informative.

We would like to express our gratitude to A.I. Baz', I.I. Gurevich, L.P. Kudrin,
V.A. Novikov, A.A. Ogloblin, I.I. Roizen, Ya.A. Smorodinskii, Yu.A. Tarasov, and
K.A. Ter-Martirosyan for discussions of the present results.

\setcounter{equation}{0}

\renewcommand{\theequation}{A.\arabic{equation}}

\section*{Appendix }

There is some methodical interest in the problem of expansion of matter which
initially occupies uniformly a spherical volume of radius $R$ at rest.
We shall assume that the temperature is ultrarelativistic.

In the limit $t\gg R$, when the one-sided expansion away from an internal weak
discontinuity $\xi=\xi_0$ has taken place, we have the general integral given by (40).
On the other side, $\xi<\xi_0$ we have a spherically symmetric simple wave.
To establish the shape of the singularity on the surface of the external weak
discontinuity, i. e., on the boundary with vacuum, let us consider the corresponding
self-similar solution (which depends only on the variable $\eta =r/t$).
Detailed analysis, which we shall omit for lack of space, leads to the following
natural-looking result
\begin{equation}\label{a1}
    s=\frac{C}{\xi_0^3t^3}\left(\frac{\zeta}{\zeta_0}\right)^3,\quad
    \ga^2=\frac12\frac{t^2}{\xi_0^2}\left(\frac{\zeta_0}{\zeta}\right)^2,
    \quad t\gg R,\quad \zeta\ll t.
\end{equation}
Here, in contrast to $\xi$, the coordinate $\zeta$ is measured from a different
bounding surface of the light signal (see also below); $\zeta_0$ is the position of
the internal weak discontinuity on the $\zeta$-scale.   In (A.1), the constants have
been chosen so as to ensure that it agrees with (40) when $\xi=\xi_0$.
For the conserved total energy and entropy, we now have $E=E'+E''$, $S=S'+S''$,
where $E'$ and $S'$ are given by (43) and $E''$ and $S''$ can be calculated by analogy,
using the solution (A.1) for the simple wave, and then integrating with respect to
$\zeta$ between zero and $\zeta_0$.   The result is
\begin{equation}\label{a2}
    S=\frac{2^{3/2}}3\pi\frac{C}{\xi_0^3}\left(1+\frac{\zeta_0}{\xi_0}\right),\quad
    E=\frac8{15}\pi k\frac{C^{4/3}}{\xi_0^5}\left(1+\frac53\frac{\zeta_0}{\xi_0}\right).
\end{equation}

To ensure that the ``asymptotic" solution (40), in general, misses the region in
which the equations of hydrodynamics are at least formally satisfied (this is
discussed below with a suitable choice of the origin of time), the sphere
$\xi =0$ must be reduced to a point.   It is clear, on the other hand, that,
in reality, this kind of singularity at $r = 0$ occurs only when the surface of the
internal weak discontinuity contracts to the origin, $t=R\sqrt{3}$.
During the same time, the external weak discontinuity will move forward to a
distance $R\sqrt{3}$.   Subsequently, the distance
\begin{equation}\label{a3}
    \zeta-\xi=\zeta_0-\xi_0=(\sqrt{3}-1)R
\end{equation}
between the singular surfaces (the spheres $\zeta =0$ and $\xi =0$) of the two
solutions to be matched undergoes no changes because both surfaces expand with
velocities strictly equal to the velocity of light.   This is the necessary third
condition which, together with (A.2), determines finally all three arbitrary
constants $C$, $\xi_0$, and $\zeta_0$.   Since
$$
k^3\frac{S^4}{E^3}=V=\frac43\pi R^3,
$$
we obtain an algebraic equation of a high (sixth) degree in the ratio $R/\xi_0$.
Numerical solution yields
\begin{equation}\label{a4}
    \begin{split}
    &\xi_0=0.52R,\quad \zeta_0=3.25R,\quad \zeta_0/\xi_0=6.3,\\
    &C=\frac{24\sqrt{2}}{125\pi}\frac{(1+5\zeta_0/3\xi_0)^3}{(1+\zeta_0/\xi_0)^5}
    R^3S.
    \end{split}
\end{equation}
Returning now to ordinary units, and using the dimensionless small combinations
$$
z'=1-\frac{r+\sqrt{3}R}{ct},\quad z''=1-\frac{r-R}{ct}
$$
for the sake of brevity, we finally obtain
\begin{equation}\label{a5}
    \begin{split}
    s=\frac{C}{(ct)^6}(z')^{-3},\quad &\ga=\frac1{\sqrt{2}}(z')^{-1},\quad
    \text{for}\quad z''>\frac{\zeta_0}{ct};\\
    s=\frac{C}{\xi_0^3\zeta_0^3}(z'')^3,\quad &\ga=\frac1{\sqrt{2}}
    \frac{\zeta_0}{\xi_0}(z'')^{-1}
    \quad\text{for}\quad 0<z''<\frac{\zeta_0}{ct};\\
    &t\gg R/c.
    \end{split}
\end{equation}
Equations (A.4) and (A.5) form the solution of our problem.

If, as a result of further expansion, the medium cools down to nonrelativistic
temperatures, the equation of state given by (30) will be violated and the
particle energy distribution will cease to vary.   We shall not consider the
details of this; the necessary derivations are similar to those leading to (47)
and (49).   We merely note that the ``weakness" of the internal discontinuity
at $\xi=\xi_0$ is reflected in the continuity of the thermodynamic and
hydrodynamic quantities (but not of their spatial derivatives).
However, to determine the particle-energy spectrum, we must transform to the
$\ga$ scale, in which case the spectral density $W(\eps)$ itself exhibits a
discontinuity.   It is not difficult to show that
\begin{equation}\label{a6}
    \frac{W_+}{W_-}=\frac{\zeta_0}{\xi_0}
\end{equation}
directly at the singularity $\eps=\eps_0=mc^2\ga(\xi_0)$ (the subscripts
$+$ and $-$ indicate the values of the functions to the right and to the left of it).
The final result is that the energy density $W(\eps)$ increases discontinuously
[see (A.4)] and thereafter decreases in accordance with the formula
$W(\eps)d\eps\propto d\eps/\eps^4$, $\eps>\eps_0$.   On the other hand,
it follows from (53) that, in the case of the problem considered in Sec.~3,
the initial geometry ensures that the distribution function $W(\eps)$ falls to
a negligible value for $\eps=\eps_0=\eps_{max}$.

\bigskip
\bigskip

\centerline{\bf Figures captions}

\bigskip

Fig.~1.   The dashed curve shows the particle spectrum in the c.m.s.
Curve $1$ refers to particles travelling in forward directions in the laboratory system.
Curve $2$ refers to particles travelling in the backward directions.   In the laboratory
system $\eps_{1max}=(6\pm2\sqrt{5})\bar{\eps}$.

\bigskip

Fig.~2. These graphs were plotted for (48) replaced by an equation in which
the proportionality factor was taken to be equal to unity.

\end{document}